\documentclass[reprint,amsmath,amssymb,nofootinbib,aps]{revtex4-2}

\usepackage{bm}% bold math
\usepackage{braket}
\usepackage{color}
\usepackage{dcolumn}
\usepackage{graphicx}
\usepackage{hyperref}
\usepackage{cleveref}
\usepackage{tabularx}
\usepackage{booktabs}
\usepackage{multirow}

\renewcommand{\section}[1]{{\par\it #1---}\ignorespaces}

\begin{document}

\title{Fixed-Point Bifurcations and Path-Information Bounds in Nonlinear Exceptional-Point Sensing}

\author{Xue-Hao Yu$^{1}$ and Cong-Feng Qiao$^{1,2}$\vspace{3mm}}
\email{corresponding author}
\vskip 2cm
\affiliation{$^1$ \small{School of Physical Sciences, University of Chinese Academy of Sciences, Beijing 100049, China\vspace{3pt}\\
$^2$ International Centre for Theoretical Physics Asia-Pacific, University of Chinese Academy of Sciences, Beijing 100190, China}}

\author{~\\\vspace{-10mm}}

\date{\today}

\begin{abstract}
Nonlinear exceptional points (NEPs) were proposed to circumvent the noise--responsivity cancellation that limits linear exceptional-point sensors. At an NEP, the stationary frequency response can be singular while the Hamiltonian Petermann factor remains finite, suggesting a genuine signal-to-noise ratio (SNR) enhancement. Yet recent model-specific analyses of nonlinear fluctuations have reached conflicting conclusions. We prove that any singular stationary frequency response requires a fixed-point bifurcation of the nonlinear dynamics. In local fluctuation regimes, the singularities in responsivity and long-time frequency uncertainty are both set by the critical relaxation rate, leading to exact cancellation in the SNR and precluding any divergent enhancement. Furthermore, we derive a path-information bound that applies across both local and nonlinear fluctuation regimes and places a finite ceiling on the long-time SNR enhancement. Notably, this upper bound is attainable in the local fluctuation regime, implying that frequency readout under such parameter settings is information-optimal.
\end{abstract}


\maketitle

\section{Introduction.}
Inferring a target perturbation $\varepsilon$ from the resulting shift $\Delta\omega$ of a resonance or oscillation frequency is a widely used measurement scheme across diverse experimental platforms, including optical microcavities \cite{VAK08,KWB+12,LCY+14}, nanomechanical resonators \cite{JKZ08,CEM+12,MGE+13}, and superconducting quantum circuits \cite{WSB+04,LVS+04,SHS+07,BGGW21}. In recent years, non-Hermitian systems operating at exceptional points (EPs) have emerged as a promising route to enhanced frequency responsivity \cite{Wie14,Wie16,Wie20a,DDS+22}. Near a generic $n$-th order EP, the frequency shift scales as $\Delta\omega\propto\varepsilon^{1/n}$, in contrast to the linear response $\Delta\omega\propto\varepsilon$ of Hermitian systems, yielding a divergent responsivity $|\partial\omega/\partial\varepsilon|\propto\varepsilon^{(1-n)/n}$ as $\varepsilon\to0$. Such responsivity enhancement has been observed across various physical implementations \cite{CKZ+17,HHW+17,HSCK19,XLKA19,MQY+20,KCE+22,WSH+26}. However, the divergent responsivity does not translate into an improved signal-to-noise ratio (SNR) in linear EP sensors \cite{Lan18,LC18,Wie20}. Near a linear EP, the increasing nonorthogonality of eigenstates causes the Petermann factor $K$ to diverge \cite{Pet79,Sie89,HW90}. The associated frequency uncertainty $\sigma_{\omega}\propto K^{1/2}\sim\varepsilon^{(1-n)/n}$ therefore scales identically with the responsivity, thereby canceling the nominal SNR enhancement \cite{WLY+20,DMA22,LS24}.

To circumvent this noise--responsivity cancellation, EP sensing has recently been extended to nonlinear systems \cite{BSJ22,SCMS22,BLLX26,DA25}. Nonlinear exceptional points (NEPs) arise from the coalescence of stationary solutions and can exhibit higher-order singularities, while the Hamiltonian remains diagonalizable and the corresponding Petermann factor stays finite \cite{BLL+23,BFL+23}. NEPs were thus expected to avoid noise amplification and deliver genuine SNR enhancement, with subsequent experiments reporting signatures of improved SNR near NEPs \cite{BLF+24,WJF+26}.

Nevertheless, recent analyses of fluctuation dynamics near NEPs have challenged these optimistic prospects. The finite Petermann factors reported in the early NEP studies were evaluated from the mean-field Hamiltonian at the stationary solutions, whereas fluctuations about a stationary solution are governed by the Bogoliubov--de Gennes (BdG) Hamiltonian of the linearized dynamics \cite{ZC25,DA26}. The frequency-uncertainty scaling near an NEP has therefore been reattributed to the BdG Petermann factor, $\sigma_{\omega}\propto K^{1/2}_{\mathrm{BdG}}$, with the associated noise amplification reported to cancel the responsivity enhancement across several models \cite{ZC25}.

\begin{figure}
    \centering
    \includegraphics[width=\linewidth]{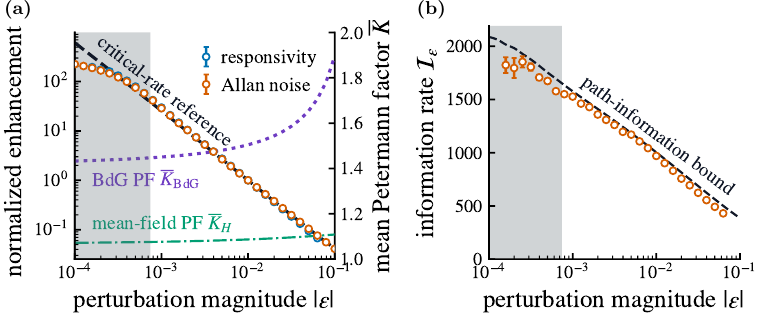}
    \caption{\textbf{Critical amplification and path-information bound.} (a) Simulated signal and Allan-noise enhancements compared with the critical-rate reference $|\mathrm{Re}\,\lambda_c|^{-2}$. The right axis displays mean-field ($K_{\mathrm{MF}}$) and BdG ($K_{\mathrm{BdG}}$) Petermann factors. (b) Simulated information rate $\mathcal{I}_\varepsilon$ saturating the critical-mode path Fisher-information rate $\dot{\mathcal{I}}_{\mathrm{path}}$. Shading marks the nonlinear regime ($\eta_2\geq1/3$).}
    \label{fig:fig-3}
\end{figure}

Across a broader class of models, however, we find that the frequency uncertainty $\sigma_{\omega}$ and $K_{\mathrm{BdG}}$ need not share the same critical behavior. Depending on the model, the noise can grow rapidly while $K_{\mathrm{BdG}}$ remains nearly flat, or remain bounded while $K_{\mathrm{BdG}}$ diverges, as shown in \cref{fig:fig-3}. More importantly, we find that a divergent responsivity is always accompanied by a corresponding divergence of the frequency uncertainty, leaving the SNR enhancement strictly bounded (see Supplemental Material). We therefore propose that the noise--responsivity cancellation is enforced by a general dynamical mechanism, rather than a divergent BdG Petermann factor.

An apparent counterexample was recently reported by Bai \textit{et al.}, in which full nonlinear Langevin simulations revealed a persistent singular response coexisting with a finite frequency uncertainty, suggesting an enhanced SNR \cite{BLX26}. They attributed the separation between response and noise to nonlinear drift confining the fluctuations. We demonstrate that the apparent separation cannot persist asymptotically, by deriving an upper bound on the SNR enhancement attainable in such nonlinear sensing schemes.

In this Letter, we establish this general dynamical mechanism by proving that any singular stationary frequency response strictly requires a fixed-point bifurcation of the nonlinear dynamics. For global $\mathrm{U}(1)$-symmetric self-oscillators, a coherent-amplitude matrix (CAM) representation eliminates the global phase, mapping rotating states to localized fixed points. A local critical-mode reduction demonstrates that responsivity and long-time frequency uncertainty share the same critical relaxation rate; this rate cancels from the SNR and yields a quantitative criterion for the breakdown of the local theory. For linearly coupled perturbations with regular noise, the residual forcing-to-noise ratio equals the path Fisher-information rate, a limit fully saturated by simple time-averaged frequency readout. Beyond local linearization, we construct a nonlinear stochastic reduction that extends the path-information bound to non-Gaussian regimes, placing a finite ceiling on the asymptotic SNR enhancement. Finally, stochastic simulations of a three-mode Kerr sensor validate the predicted critical scaling and nonlinear crossover, confirming that the extracted frequency information strictly obeys our path-information bound.

\section{Coherent Amplitude Dynamics.} 
Previous studies of nonlinear exceptional-point sensing \cite{BLL+23,BFL+23,DA25,DA26,BLF+24,WJF+26,ZC25,BLX26} are formulated in terms of the $c$-number Langevin equation
\begin{equation}\label{eqn:langevin} 
\dot{\bm{\alpha}}=-i\bm{H}(\bm{\alpha})\bm{\alpha}+\bm{\xi}(t),
\end{equation}
where $\bm{\alpha}=(\alpha_{1},\alpha_{2},\ldots,\alpha_{n})^T$ denotes the complex mode amplitudes, $\bm H(\bm\alpha)$ is an effective nonlinear Hamiltonian, and $\bm\xi(t)$ is a zero-mean Gaussian white noise satisfying $\langle\bm{\xi}(t)\bm{\xi}^{\dagger}(t')\rangle=\bm{D}(\bm{\alpha})\delta(t-t')$. In the noise-free limit, a steady self-oscillation takes the form of a stationary rotating solution $\bm{\alpha}(t)=\bm{\alpha}_{0}e^{-i\omega t}$, and the perturbation-induced shift of its rotation frequency serves as the sensing signal.

In the presence of noise, stochastic diffusion broadens the deterministic rotating solution into an annular distribution, where amplitude and relative-phase fluctuations are obscured by the unconstrained motion of the global phase. To isolate the phase-invariant dynamics, we adopt a coherent-amplitude matrix (CAM) representation that quotients the state space by the continuous symmetry group. By removing motion along the symmetry orbit, the CAM mapping transforms the rotating orbit into a fixed point and its noise-broadened distribution into a localized steady state in CAM space, enabling the response, stability, and dominant fluctuation directions to be analyzed about a fixed reference. Since the NEP models considered in previous studies \cite{BLL+23,BFL+23,BLF+24,ZC25,DA25,DA26,WJF+26,BLX26} share a global $\mathrm{U}(1)$ phase symmetry, we focus on the corresponding $\mathrm{U}(1)$-invariant representation, with extensions to other continuous symmetries detailed in the Supplemental Material.

\begin{figure}
    \centering
    \includegraphics[width=\linewidth]{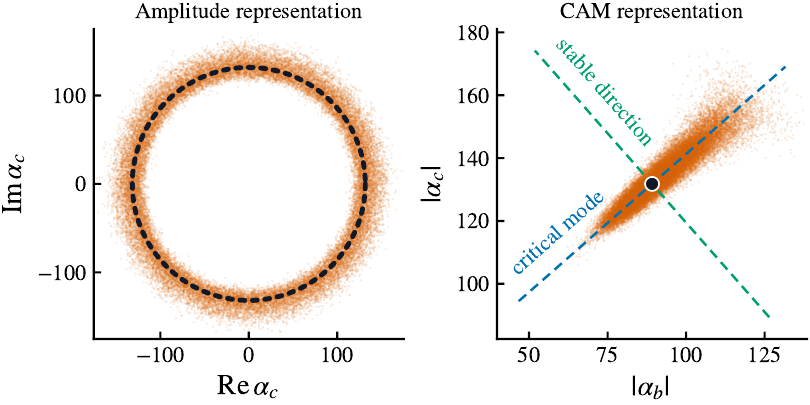}
    \caption{\textbf{Coherent-amplitude matrix (CAM) representation.} Left: Annular steady-state distribution in complex-amplitude space; the dotted line shows the noise-free rotating orbit. Right: Localized distribution under the CAM mapping $\bm{R}=\bm{\alpha}\bm{\alpha}^{\dagger}$, with the noise-free fixed point (dot). Dashed lines indicate the critical mode and orthogonal stable direction of the CAM Jacobian.}
    \label{fig:fig-1}
\end{figure}

For each stochastic trajectory, we define the coherent-amplitude matrix $\bm{R}=\bm{\alpha}\bm{\alpha}^\dagger$. Since both the Hamiltonian and the diffusion matrix are phase-invariant, applying It\^o calculus \cite{Gar09} to evaluate the deterministic drift yields
\begin{equation}\label{eqn:R-dynamics}
\dot{\bm{R}}=\mathcal{L}(\bm{R})=-i\bm{H}(\bm{R})\bm{R}+i\bm{R}\bm{H}^\dagger(\bm{R})+\bm{D}(\bm{R}).
\end{equation}
The diffusion matrix $\bm{D}(\bm{R})$ thus contributes directly to the CAM flow. For a noise-free rotating solution $\bm{\alpha}(t)=\bm{\alpha}_{0}e^{-i\omega t}$, the global phase cancels exactly in $\bm{R}(t)$, yielding a fixed point $\mathcal{L}(\bm{R}_{0})=0$, as illustrated in \cref{fig:fig-1}. The response and stability can therefore be analyzed through the CAM fixed points.

Although the global phase is eliminated from $\bm{R}$, the oscillation frequency remains directly accessible from the CAM dynamics. In the noise-free limit, a stationary rotating solution $\bm{\alpha}(t)=\bm{\alpha}_{0}e^{-i\omega t}$ defines a rank-one CAM fixed point $\bm{R}_{0}=\bm{\alpha}_{0}\bm{\alpha}_{0}^{\dagger}$. The oscillation frequency $\omega$ is given exactly by the generalized Rayleigh functional
\begin{equation}\label{eq:freq-fixed-point}
\omega=\mathrm{Re}\,\frac{\mathrm{Tr}\!\left[\bm{H}(\bm{R}_{0})\bm{R}_{0}\right]}{\mathrm{Tr}(\bm{R}_{0})}.
\end{equation}
More generally, in the presence of fluctuations, the steady-state oscillation appears as a spectral line of finite linewidth, whose peak defines the carrier frequency. The corresponding CAM fixed point is no longer rank-one and cannot be reduced to a single rotating vector $\bm{\alpha}(t)=\bm{\alpha}_{0}e^{-i\omega t}$. Within the local Ornstein--Uhlenbeck (OU) approximation, the carrier frequency can nevertheless be recovered from a general CAM fixed point via \cref{eq:freq-fixed-point} (see Supplemental Material), with higher-order corrections from nonlinear stochastic dynamics addressed later in this Letter.

\section{Fixed-Point Bifurcations.}
We consider a dispersive sensing scheme in which a target perturbation $\varepsilon$ modifies the system dynamics and can thus be inferred from the resulting frequency shift. As the target perturbation $\varepsilon$ is varied, the CAM fixed points form a stable branch $\bm{R}_{c}(\varepsilon)$ satisfying $\mathcal L(\bm R_c(\varepsilon),\varepsilon)=0$. To characterize the response along this branch, we introduce the Jacobian of the CAM flow $\mathcal{J}=D_{\bm {R}}\mathcal{L}(\bm{R}_{c},\varepsilon)$ and the parameter forcing matrix, $\mathcal{F}=\partial_\varepsilon\mathcal{L}(\bm{R}_{c},\varepsilon)$. Differentiating the steady-state condition with respect to $\varepsilon$ yields the fixed-point response equation $\mathcal{J}\!\left(\mathrm{d}\bm{R}_{c}/\mathrm{d}\varepsilon\right)+\mathcal{F}=0$, which determines a regular response $\mathrm{d}\bm{R}_{c}/\mathrm{d}\varepsilon=-\mathcal{J}^{-1}\mathcal{F}$ provided $\mathcal{J}$ is invertible. A singular response thus requires $\mathcal{J}$ to be non-invertible, corresponding to a fixed-point bifurcation. We focus on the canonical case where a single real eigenvalue of $\mathcal{J}$ approaches zero, in which case the singular response and dominant fluctuations are then entirely governed by the corresponding critical mode.

Along the stable branch, let $\lambda_{c}<0$ denote the critical eigenvalue and let $\bm{\Phi}_{c}$ and $\bm{\Psi}_{c}$ be its right and left eigenmatrices, normalized such that $\langle\bm{\Psi}_{c},\bm{\Phi}_{c}\rangle=1$ under the Hilbert--Schmidt inner product. To relate the target perturbation to the observable frequency $\omega(\bm{R},\varepsilon)$ defined by \cref{eq:freq-fixed-point}, we define two physical coupling coefficients: $F_{c}=\langle\bm{\Psi}_{c},\mathcal{F}\rangle$, which quantifies the effective drive injected into the critical mode, and $g_{c}=\left\langle\nabla_{\bm{R}}\,\omega(\bm{R}_c,\varepsilon),\bm{\Phi}_{c}\right\rangle$, which represents the frequency shift per unit displacement along the critical mode. When both couplings are nonzero, projecting the fixed-point response equation onto the critical mode yields the singular frequency responsivity,
\begin{equation}\label{eq:critical-response}
\mu_{\omega}:=\frac{\mathrm{d}\omega}{\mathrm{d}\varepsilon}
=-\frac{g_{c}F_{c}}{\lambda_{c}}+O(1).
\end{equation}
The divergent responsivity is therefore set by the vanishing relaxation rate, with $|\mu_{\omega}|\propto|\lambda_{c}|^{-1}$.

\begin{figure}
    \centering
    \includegraphics[width=\linewidth]{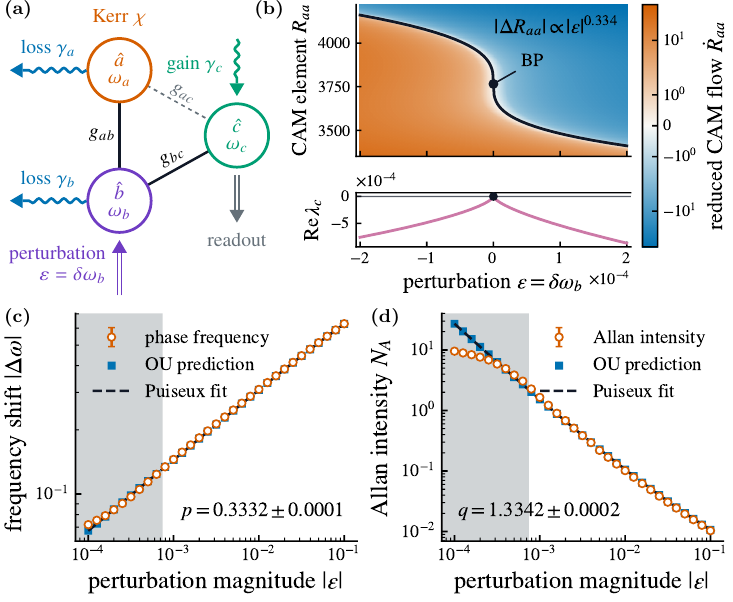}
    \caption{\textbf{Cubic critical response and noise in a three-mode Kerr sensor.} (a) System schematic with Kerr mode $\hat{a}$, perturbed mode $\hat{b}$, and active readout mode $\hat{c}$. (b) Reduced CAM flow and vanishing critical eigenvalue $\lambda_c$ near the cusp bifurcation point (BP). (c) Carrier-frequency shift showing $|\Delta\omega|\propto\varepsilon^{1/3}$ critical scaling compared with the CAM prediction. (d) Allan intensity showing $N_{A}\propto\varepsilon^{-4/3}$ scaling. Light-gray shading marks the nonlinear regime ($\eta_2\geq1/3$).}
    \label{fig:fig-2}
\end{figure}

To determine the asymptotic scaling of $\lambda_{c}$ with $\varepsilon$, we examine the nonlinear dynamics near the bifurcation point $\bm{R}_{0}=\bm{R}_{c}(0)$, where noncritical modes relax rapidly and are slaved to the critical mode \cite{Hak13}. By the center-manifold theorem \cite{Kuz13}, the dynamics collapses onto a one-dimensional center manifold tangent to $\bm{\Phi}_{c}$, which can be parameterized locally by a scalar coordinate $x$ as $\bm{R}_{c}(\varepsilon)=\bm{R}_{0}+x(\varepsilon)\bm{\Phi}_{c}+O(x^{2},\varepsilon)$. Projecting the CAM flow onto $\bm{\Psi}_{c}$ yields the reduced scalar dynamics with leading normal form
\begin{equation}\label{eq:normalform}
\dot{x}=f(x,\varepsilon)=F_{c}\varepsilon-d_{n}x^{n}+\cdots, \qquad n\geq2.
\end{equation}
where $-d_{n}x^n$ represents the leading restoring drift of the critical mode, with the expansion coefficients determined by the CAM flow derivatives (see Supplemental Material). Along a stable real branch, the stationary condition $f(x_{c},\varepsilon)=0$ gives $x_{c}\propto|\varepsilon|^{1/n}$ and hence $|\lambda_{c}|=\left|\partial_{x}f(x_{c},\varepsilon)\right|\propto|\varepsilon|^{(n-1)/n}$. Thus, the carrier frequency $\omega(\varepsilon)$ inherits the fractional-power dependence of the critical coordinate $x(\varepsilon)$, and the vanishing restoring rate produces a divergent responsivity, as illustrated in \cref{fig:fig-2}.

\section{Local Signal-to-Noise Cancellation.}
The critical mode governing the singular frequency response also dominates the local fluctuations near the bifurcation. For fixed $\varepsilon$, we write $\delta x(t)=x(t)-x_{c}(\varepsilon)$ for the fluctuation about the stable fixed point of the reduced dynamics. Linearizing $f(x,\varepsilon)$ about $x_{c}$ and including the stochastic force projected onto the critical mode gives the Ornstein--Uhlenbeck process
\begin{equation}\label{eq:critical-OU}
\delta\dot{x}=\lambda_{c}\delta x+\xi_{c}(t),
\quad
\langle\xi_{c}(t)\xi_{c}(t')\rangle=\Gamma_{c}\delta(t-t'),
\end{equation}
where $\Gamma_{c}$ denotes the noise intensity injected into the critical coordinate (see Supplemental Material).

Solving \cref{eq:critical-OU} in the stationary regime gives the fluctuation variance $\sigma_{x}^{2}:=\langle\delta x^{2}\rangle=\Gamma_{c}/(2|\lambda_{c}|)$, while the fluctuations relax over the correlation time $\tau_{c}=|\lambda_{c}|^{-1}$. In turn, critical-mode fluctuations map to carrier-frequency fluctuations via $\delta\omega(t)=g_{c}\delta x(t)$. The uncertainty of the carrier frequency over an averaging time $\tau$ is quantified by the Allan deviation $\sigma_{A}(\tau)$, which follows a $\tau^{-1/2}$ dependence for $\tau\gg\tau_{c}$. We denote the long-time Allan intensity by
\begin{equation}\label{eq:critical-frequency-noise}
N_{A}:=\lim_{\tau\to\infty}\tau\sigma_{A}^{2}(\tau)=S_{\omega}(0)=\frac{|g_{c}|^{2}\Gamma_{c}}{|\lambda_{c}|^{2}}.
\end{equation}
For an observation time $T\gg\tau_{c}$, the SNR for detecting a parameter variation $\Delta\varepsilon$ is defined as the ratio of the induced carrier-frequency shift to the Allan deviation,
\begin{equation}\label{eq:local-cancellation}
\mathrm{SNR}:=\frac{|\Delta\omega|}{\sigma_{A}(T)}\sim\frac{|\mu_{\omega}\Delta\varepsilon|}{\sqrt{N_{A}/T}}\sim\sqrt{T}|\Delta\varepsilon|\cdot\frac{|F_{c}|}{\sqrt{\Gamma_{c}}}.
\end{equation}

Within the local OU regime, the vanishing restoring rate of the critical CAM mode amplifies both the responsivity and the frequency-noise amplitude by the same factor $|\lambda_{c}|^{-1}$, which therefore cancels from the SNR. The underlying amplification mechanism does not rely on eigenbasis nonorthogonality in either the mean-field Hamiltonian $\bm{H}(\bm{R}_{c})$ or the Bogoliubov--de Gennes (BdG) fluctuation operator. Nonorthogonality of the CAM Jacobian eigenmodes can nevertheless provide additional noise amplification through the projected intensity $\Gamma_{c}$. In particular, the projected noise intensity factorizes as $\Gamma_{c}=K_{c}D_{0}\hat{d}_{c}$, where $K_{c}$ is the dynamical Petermann factor of the critical CAM mode, $D_{0}$ is the overall noise strength, and $\hat{d}_{c}$ describes the diffusion projected onto that mode (see Appendix B). When the CAM Jacobian $\mathcal{J}$ approaches an exceptional point, $K_{c}$ can diverge and provide additional noise amplification, which is responsible for the excess BdG noise reported in Ref.~\cite{ZC25}.

The local cancellation in \cref{eq:local-cancellation} relies on linearizing the reduced drift. As the bifurcation is approached $\lambda_{c}\to0$, the growing fluctuation variance $\sigma_{x}$, causes nonlinear drift terms to become dominant. We characterize the validity threshold of the local OU theory by comparing the leading quadratic term against the linear restoring term at $\lambda_{c}\delta x$ at $|\delta x|\sim\sigma_{x}$. For the quadratic correction, we define $\eta_{2}=|f''(x_{c})|\sigma_{x}/\left(2|\lambda_{c}|\right)$. The local OU approximation breaks down when $\eta_2=O(1)$, marking the crossover to nonlinear stochastic dynamics, as illustrated in \cref{fig:fig-2}.

\section{Nonlinear Fluctuations and Information Bound.}
Beyond the local OU regime, we retain the full nonlinear drift $f(x,\varepsilon)$ in \cref{eq:normalform}  together with the stochastic drive projected onto the critical coordinate. The critical dynamics is then governed by the reduced one-dimensional diffusion
\begin{equation}\label{eq:nonlinear-reduction}
\mathrm{d}x_{t}=f(x_{t},\varepsilon)\,\mathrm{d}t+\sqrt{q(x_{t},\varepsilon)}\,\mathrm{d}W_{t},
\end{equation}
where, $q(x,\varepsilon)$ denotes the projected diffusion matrix (see Supplemental Material). This reduced description fully captures non-Gaussian fluctuations and their nonperturbative contributions to frequency noise.  

Under stationary and ergodic conditions, \cref{eq:nonlinear-reduction} defines a stationary density $\pi_\varepsilon(x)$. For a stationary path $\{x_t\}$, the time-averaged carrier frequency $\widehat{\omega}_T$ averaged over an observation time $T$ estimates the ensemble average $\overline{\omega}(\varepsilon):=\langle\omega(x_t)\rangle_{\pi_\varepsilon}$, with the long-time Allan intensity
\begin{equation}
\begin{aligned}
N_{A}&=2\int_{0}^{\infty}{\operatorname{Cov}_{\pi_{\varepsilon}}[\omega(x_{t}),\omega(x_{0})]\,\mathrm{d}t}\\
&=\lim_{T\to\infty}{T\,\operatorname{Var}_{\varepsilon}(\widehat{\omega}_{T})}.
\end{aligned}
\end{equation}
Since the full CAM dynamics has been reduced to the one-dimensional center manifold associated with the critical mode, both $\pi_{\varepsilon}$ and $N_{A}$ admit exact quadrature representations, enabling exact evaluation without solving a time-dependent Fokker--Planck equation (see Supplemental Material).

Crucially, no frequency estimator constructed from the reduced stochastic path, including $\widehat{\omega}_{T}$, can contain more information about $\varepsilon$ than the complete trajectory. Assuming the parameter perturbation enters through the drift, Girsanov's theorem gives the asymptotic Fisher information rate of the reduced path \cite{Gir60,LS00,DKPP16},
\begin{equation}\label{eq:path-FI-rate}
\dot{\mathcal{I}}_{\mathrm{path}}:=\lim_{T\to\infty}\frac{\mathbb{E}_{\varepsilon}[\mathcal{S}_{T}^{2}]}{T}=\left\langle\frac{|\partial_{\varepsilon}f(x,\varepsilon)|^2}{q(x)}\right\rangle_{\pi_{\varepsilon}},
\end{equation}
where $\mathcal{S}_{T}$ is the path score function. Applying the Cauchy--Schwarz inequality to the score identity $\partial_{\varepsilon}\mathbb{E}_{\varepsilon}[\widehat{\omega}_{T}]=\operatorname{Cov}_{\varepsilon}(\widehat{\omega}_{T},\mathcal{S}_{T})$ establishes a fundamental information bound on the local asymptotic SNR $\Delta\varepsilon\to0$,
\begin{equation}\label{eq:path-information-bound}
\lim_{T\to\infty}\frac{\mathrm{SNR}^{2}}{T|\Delta\varepsilon|^{2}}=\frac{|\mathrm{d}\overline{\omega}/\mathrm{d}\varepsilon|^{2}}{N_{A}}\leq\dot{\mathcal{I}}_{\mathrm{path}}.
\end{equation}
Since $\dot{\mathcal I}_{\mathrm{path}}$ remains finite across the bifurcation (see Supplemental Material), the SNR enhancement is strictly bounded even when the frequency responsivity diverges, extending the signal-to-noise constraint in \cref{eq:local-cancellation} to the fully nonlinear regime.

In most measurement setups, a weak parameter perturbation $\varepsilon$ couples linearly to the Hamiltonian, $\bm{H}(\bm{\alpha},\varepsilon)=\bm{H}_{0}(\bm{\alpha})+\varepsilon\bm{V}(\bm{\alpha})$, so the critical dynamics simplifies to $f(x,\varepsilon)=f_{0}(x)+F_{c}\varepsilon$ with a constant noise floor $q(x)=\Gamma_{c}$. The path information rate is then independent of the nonlinear drift $f_{0}(x)$, directly giving $\dot{\mathcal{I}}_{\mathrm{path}}=|F_{c}|^{2}/\Gamma_{c}$. Consequently, the fundamental information bounds in \cref{eq:path-information-bound} reduce to
\begin{equation}\label{eq:nonlinear-information-ceiling}
N_{A}\geq\frac{\Gamma_{c}}{|F_{c}|^{2}}\left|\frac{\mathrm{d}\overline{\omega}}{\mathrm{d}\varepsilon}\right|^{2},
\quad
\lim_{T\to\infty}\frac{\mathrm{SNR}^{2}}{T|\Delta\varepsilon|^{2}}\leq\frac{|F_{c}|^{2}}{\Gamma_{c}}.
\end{equation}
The signal--noise cancellation in \cref{eq:local-cancellation} saturates this bound in the OU regime, establishing that linear signal cancellation reflects the physical equality limit of path information rather than a truncation artifact. Saturation requires the scalar frequency readout to retain all Fisher information of the trajectory, which is generally violated under nonlinear drift or readout.

\section{Discussion.}
Our results establish a general dynamical framework for singular frequency response in nonlinear sensors: when a fixed-point branch develops a vanishing relaxation rate, any perturbation driving the critical mode produces a divergent frequency response. Crucially, this framework operates on full nonlinear physical dynamics rather than linearized or mean-field Hamiltonian spectra. Nonlinear exceptional points are thus special manifestations of a broader class of fixed-point bifurcations, meaning that direct bifurcation analysis of the CAM flow can uncover critical sensing points hidden to standard Hamiltonian defectiveness criteria \cite{Wie14,ZC25,DA26}. Our framework thus unifies and extends earlier model-specific connections between Hamiltonian EPs and nonlinear bifurcations \cite{ZWC19,UTN25}, demonstrations of singular responses without Hamiltonian defectiveness \cite{ZRX+23,FBG+25}, and catastrophe-theoretic classifications of nonlinear eigenproblems~\cite{KWA+26}.

The critical mode dictates both singular response and frequency fluctuations. In the local OU regime, the critical factor $|\lambda_c|^{-1}$ cancels between responsivity and noise, reducing the asymptotic sensing capacity to the projected information rate $|F_c|^2/\Gamma_c = \dot{\mathcal{I}}_{\mathrm{path}}$. The true physical resource for critical sensing is thus this projected information rate rather than divergent responsivity alone; genuine SNR enhancements require $|F_c|^2/\Gamma_c$ to exceed that of a reference operating point. While dynamical Petermann factors modulate $\Gamma_c$ and adjust attainable finite enhancement, they do not alter the underlying critical amplification shared by signal and noise. This local cancellation and its nonlinear saturation fully reconcile the bounded-SNR results of Refs.\,\cite{ZC25,DA26}. Beyond the local regime, nonlinear drift confines spatial fluctuations \cite{BLX26} while intrinsically modifying the carrier response. Our path-information bound accounts for both effects, ruling out unbounded asymptotic enhancement while preserving finite, practical SNR enhancements at nonzero signal bias or over finite measurement windows \cite{BLX26,BLF+24,WJF+26}.

Importantly, the simple time-averaged frequency estimator fully saturates the trajectory's Fisher information in the local OU regime, rendering complicated post-processing redundant for long-time sensitivity. For sensor design, this reveals that operating arbitrarily near the bifurcation yields no intrinsic information gain once signal and noise share critical scaling; instead, critical slowing down merely prolongs the time needed to acquire independent statistical samples. The local OU regime thus defines the statistically efficient operating window. Beyond it, non-Gaussian fluctuations prevent the mean frequency from harvesting all path information. Optimal sensor bias should therefore maximize $|F_c|^2/\Gamma_c$ subject to bandwidth and detector noise constraints, rather than blindly maximizing responsivity.

Beyond frequency-shift detection, our framework provides a unified strategy for nonlinear critical sensing: candidate operating points are identified via reduced-flow bifurcations and evaluated through critical-mode projections of perturbation, noise, and readout. While the $\mathrm{U}(1)$ CAM suits self-oscillators, analogous symmetry-invariant representations apply to systems with other continuous symmetries. Extending this approach to driven-dissipative probe sensors \cite{PR22,DMP+23} will require incorporating scattering dynamics and detector noise. For systems with multiple critical modes or time-dependent attractors, higher-dimensional stochastic reductions can connect our fixed-point theory to generalized exceptional dynamics on non-stationary attractors \cite{WFH+25}. Together, these directions establish a systematic foundation for comparing critical sensing mechanisms across diverse nonlinear platforms.

%%%%%%%%%%%%%%%%%%%%%%%%%%%%%%%%%%%%%%%%%%%%%%%%%%%%%%%%%%%%%%%%%%%%%
\vspace{0.5cm} 
\textit{Acknowledgments}---This work was supported in part by the National Natural Science Foundation of China(NSFC) under the Grants 12475087 and 12235008. 
%%%%%%%%%%%%%%%%%%%%%%%%%%%%%%%%%%%%%%%%%%%%%%%%%%%%%%%%%%%%%%%%%%%%

\bibliography{ref.bib}% Produces the bibliography via BibTeX.

\end{document}